\documentclass[fleqn,10pt,twocolumn]{AROB-ISBC-SWARM22}

\usepackage{amsmath}
\usepackage{graphicx}

\title{A study on the ephemeral nature of knowledge shared within multiagent systems}

\author{Sanjay Sarma O V${}^{1\dagger}$ and Ramviyas Parasuraman${}^{2}$ and Ramana Pidaparti${}^{1}$ }
\speaker{Sanjay Sarma O V}

\affils{${}^{1}$College of Engineering, University of Georgia, Athens, GA 30602, USA\\
${}^{2}$School of Computing, University of Georgia, Athens, GA 30602, USA\\
(Corresponding author e-mail: ramviyas@uga.edu)\\
}
\abstract{%
Achieving knowledge sharing within an artificial swarm system could lead to significant development in autonomous multiagent and robotic systems research and realize collective intelligence.
However, this is difficult to achieve since there is no generic framework to transfer skills between agents other than a query-response-based approach.
Moreover, natural living systems have a "forgetfulness" property for everything they learn. Analyzing such ephemeral nature (temporal memory properties of new knowledge gained) in artificial systems has never been studied in the literature.
We propose a behavior tree-based framework to realize a query-response mechanism for transferring skills encoded as the condition-action control sub-flow of that portion of the knowledge between agents to fill this gap. 
We simulate a multiagent group with different initial knowledge on a foraging mission. While performing basic operations, each robot queries other robots to respond to an unknown condition. The responding robot shares the control actions by sharing a portion of the behavior tree that addresses the queries. 
Specifically, we investigate the ephemeral nature of the new knowledge gained through such a framework, where the knowledge gained by the agent is either limited due to memory or is forgotten over time. Our investigations show that knowledge grows proportionally with the duration of remembrance, which is trivial. However, we found minimal impact on knowledge growth due to memory. We compare these cases against a baseline that involved full knowledge pre-coded on all agents. We found that knowledge-sharing strived to match the baseline condition by sharing and achieving knowledge growth as a collective system. 
}


\begin{document}

\maketitle

%-----------------------------------------------------------------------

\section{Introduction}
%We present the results of this extensive analysis, with insights from the study that will have immense significance in understanding and realizing collective intelligence in artificial systems.
Collective intelligence achieved through sharing vital information and knowledge between members of the same species is a common trait in nature and animals. These include sharing travel information, food opportunities, preparation, predator locations, vocalizations, hunting, etc. This information is crucial for the survival of the animals, and the species overall, which is widely observed in nature \cite{Caro1992IsAnimals,Schweinfurth2020TheNorvegicus}. Developing similar knowledge-sharing strategies in multiagent and swarm systems has advantages in improving coordination, and accomplishing complex missions \cite{williams2004learning,Jones2018EvolvingRobotics,Iovino2020AAI}.

%For example, The Norway rats (\textit{Rattus norvegicus}) gain knowledge through social learning, helping them understand toxic diets and hunting techniques \cite{Schweinfurth2020TheNorvegicus}. On the other hand, Songbirds evolved to demonstrate song learning behavior similar to language acquisition in humans, which is used to mark their territory and attract females \cite{BarrosCLearning}. Another example includes that of bottle-nose dolphins \cite{Wild2020IntegratingInnovation} sharing foraging knowledge with their offspring through vertical social transmission observed in the Indo-Pacific region.

%Developing similar knowledge-sharing strategies in multiagent and swarm systems has advantages in improving coordination accomplishing complex missions. Jones et al. \cite{Jones2018EvolvingRobotics} developed an evolutionary algorithm on a kilobot swarm applied to behavior trees of the individual robots. The algorithm evolves an optimal behavior tree by the end of the simulations for a foraging problem that is collectively utilized. Ramiro et al. \cite{Agis2020AnGames} propose an event-driven behavior tree framework that uses request handlers to execute coordinated behaviors based on the communication messages exchanged. Other research also points to sharing policies between multiple agents for better coordination in complex missions.

While there have been extensive studies on transferring knowledge between living beings (non-human animals) in ecology, developing and applying similar knowledge-sharing strategies in artificial multiagent and swarm systems is very limited \cite{williams2004learning}. Researchers in the past used learning through clarification mechanism \cite{chernova2008teaching,chernova2009confidence} or action selection mechanisms \cite{erdogan2011action} to gain new knowledge of an artificial agent such as a robot. However, we find a lack of a generic framework for knowledge representation and strategies for transferring knowledge across agents from these works. 
Moreover, the ephemeral properties of such new knowledge gained in an artificial swarm system have not been studied to analyze the collective system performance in the literature.

Along these lines, we contribute two first-of-kind innovations. First, we present a novel representation of knowledge through the concept of behavior trees \cite{colledanchise2018behavior} borrowed from the game development community and create a framework for sharing knowledge between multiple agents encoded by implementing an efficient query, respond and update (learn) mechanism (similar to the learning from clarification \cite{chernova2008teaching}) encoded within a swarm agent behavior. 
Secondly, we study the ephemeral nature (short-term memory) of the shared knowledge between swarm agents by conducting extensive numerical analysis in various scenarios for a simulated foraging task. Through these innovations, we analyze how the memory of shared knowledge affects overall performance.

\section{Study Objectives}
We use behavior trees for state-action transition control of individual agents, which encodes the heterogeneity in the knowledge \cite{ov2020impact}. Behavior trees (BT) are structures designed for autonomous decision-making in agents. They were first developed to define the Non-Player Character (NPC) \cite{Robertson2015BuildingGames} control structures in gaming and have also found widespread application in robotics, and artificial intelligence \cite{Colledanchise2019LearningAgents,Iovino2020AAI}.

\begin{figure}[b]
\begin{center}
\includegraphics[width=\linewidth]{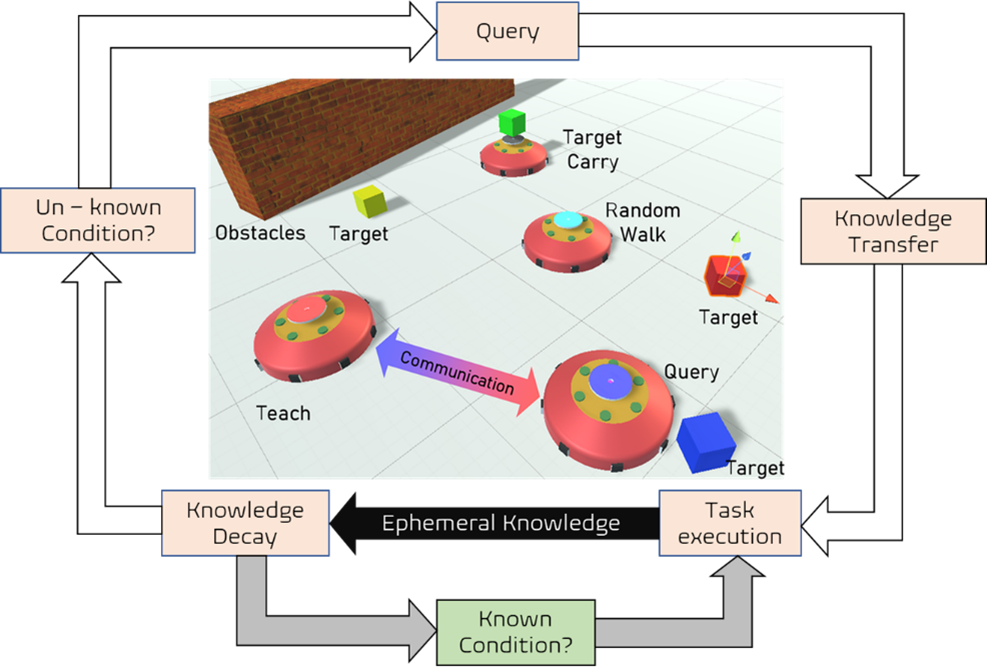}
\caption{\label{Fig:ELC} A snapshot of the simulator used in the current study and the ephemeral learning cycle.}
\end{center}
%\vspace{-4mm}
\end{figure}

To perform this study, we developed a graphical simulator for a swarm foraging problem with robots searching for various colored food targets (Red, Green, Yellow, and Blue) to demonstrate the knowledge transfer across agents. Each agent in the simulator is initialized with a portion of knowledge in a behavior tree and is allowed to communicate with other agents through queries and answers. An agent facing an unknown set of conditions broadcasts a question and awaits a response. On the other hand, an agent receiving the query can respond to share its knowledge through an encoded BT. The querying agent merges this information in its knowledge and continues to perform the required action for the conditions faced. In our work, we encode an agent's knowledge through a BT string, which facilitates knowledge sharing.

Through these simulations, we investigate the knowledge propagation across the group, the performance of the group in target collection, the effect of the memory duration, and the effect of memory size (measured in the number of subtrees) on the group's performance in the case of ephemeral learning. This is compared against baseline performance that involved no learning or teaching.

\iffalse
\begin{figure*}[t]
\includegraphics[width=0.25\textwidth]{TargetCountGraph_MSize.png}
\includegraphics[width=0.25\textwidth]{TargetCountGraph_MTime.png}
\includegraphics[width=0.25\textwidth]{KnowledgePerc_MSize.png}
\includegraphics[width=0.25\textwidth]{KnowledgePerc_MTime.png}
\end{figure*}
\fi

\vspace{-5pt}
\section{Simulation Results}
We conducted simulations to study ephemeral learning on a foraging simulator, as shown in Fig. \ref{Fig:ELC}.
Below, we present the results of a specific study on understanding the ephemeral nature of the knowledge shared between swarm agents. Here, an agent can only remember the knowledge received from other agents for a specific duration and forgets that ability after the timeout. This assimilates the knowledge decay behavior observed in animals and human society to collect information from various sources with varying significance. 

We simulated swarms of homogeneous type agents, with different groups having different abilities. 
%An outline of the ephemeral learning mechanism is presented in Fig. \ref{Fig:ELC}. 
We used a variety of robots that have different starting knowledge states. An \textit{I} robot is ignorant and has no knowledge of any of the four colored targets, and an \textit{M} robot, on the contrary, has knowledge about all the targets. \textit{R, G, Y, and B} robots handle only their corresponding colors (For e.g., an \textit{R} Robot handles only Red targets). 
A summary of simulation parameters used is presented in Table \ref{Table:Parameters}, which shows the studied memory duration properties (T1K -- T20K, where 20K iterations are the entire duration of one trial) and memory storage properties (the robot can store knowledge of up to X targets in the MX case).

\begin{table}[t]
\caption{\label{Table:Parameters}Parameters used in the simulation cases.}
 \resizebox{\columnwidth}{!}{%
\begin{tabular}{|c|c|}
\hline\hline
\textbf{Parameter}                  & \textbf{Value}           \\ \hline
Total Targets (R, G, Y, B)           & (25, 25, 25, 25)         \\
Robots (I, M, R, G, Y, B)            & (45, 5, 0, 0, 0, 0)      \\
Memory Duration (Iters.)  & 1000, 2000, 5000, 10000, 20000 \\
Memory Size (Sub-trees)  & 1, 2, 3, and 4 \\
Iterations (max)                     & 20000                    \\
%Trials                                                    & 10                       \\
\hline
\end{tabular}
}%
%\vspace{-4mm}
\end{table}

From the simulation data, we computed the mean for the total number of targets captured over 10 trials and the percentage of the knowledge possessed by the group over iterations according to Eq.~\eqref{Eq:eq1}.  
%\vspace{-3pt}
\begin{equation}
%\begin{array}{rcl}
Knowledge\% = \frac{R_k+G_k+Y_k+B_k}{Max. Possible Knowledge}\times 100
%\end{array}
\label{Eq:eq1}
\end{equation}

Here, $R_k,G_k,Y_k,B_k$ are the counts of the numbers of robots that possess knowledge about the red, green, yellow, and blue targets, respectively. And $Max.PossibleKnowledge$ is the maximum possible knowledge when all the robots know to handle all the targets. (i.e., $50\times4=200$). For the ephemeral cases, the initial knowledge is 10\% since there are only 5 \textit{M} robots and 45 \textit{I} robots. We compare these cases against a baseline (BL) that has 50 \textit{M} robots with knowledge of 100\%. 
%The knowledge percent values obtained from equation \ref{Eq:eq1} computed for BL stands at 100\%. 

\begin{figure}[h]
%\vspace{-4mm}
\begin{center}
\includegraphics[width=1\columnwidth]{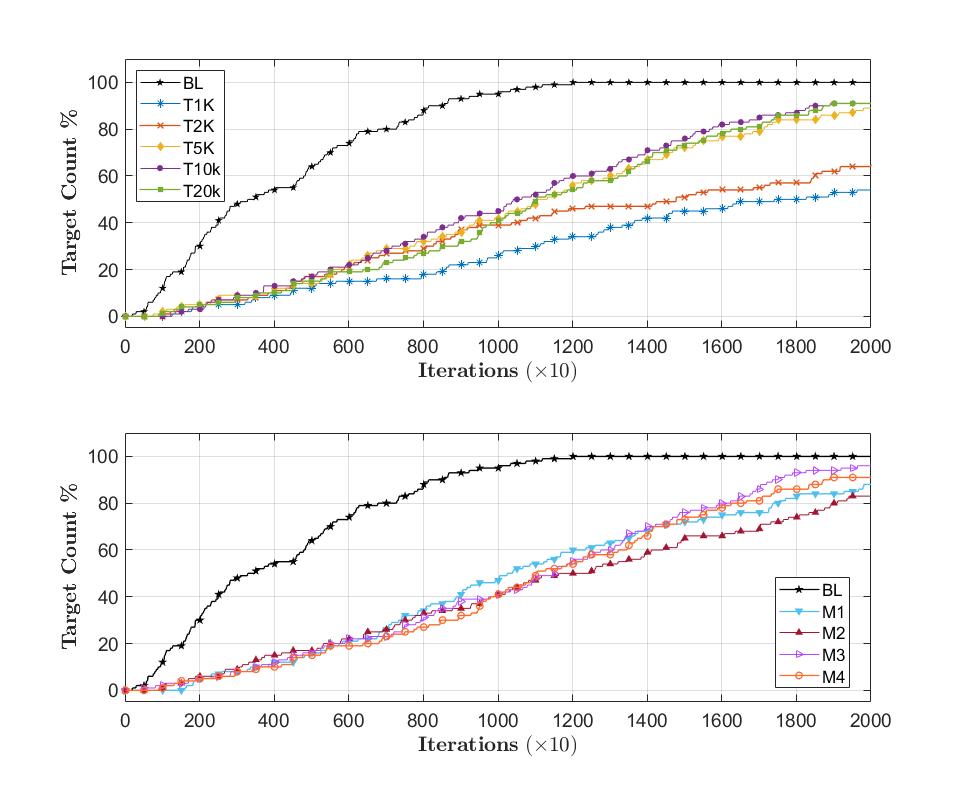}
\caption{\label{Fig:Target} Rates of target collection for different analysis types.}
\end{center}
\end{figure}

\begin{figure}[h]
\vspace{-10pt}
\begin{center}
\includegraphics[width=1\columnwidth]{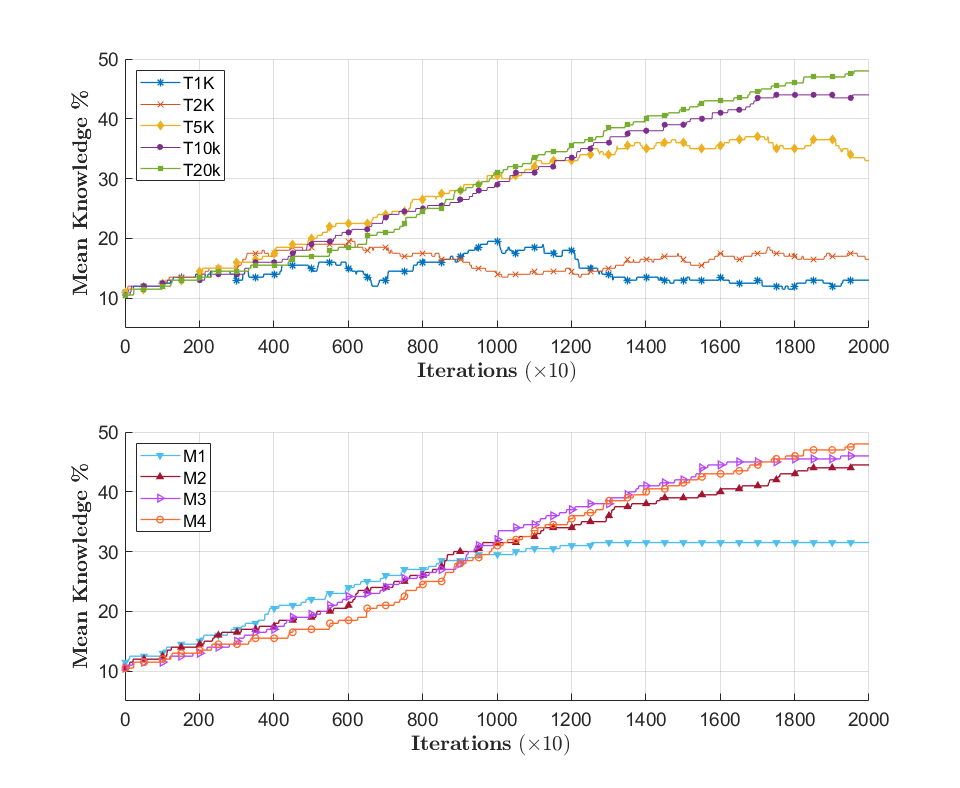}
\caption{\label{Fig:Knowledge} Knowledge gain vs. time for different analysis types.  }
\end{center}
%\vspace{-20pt}
\end{figure}

Figs.~\ref{Fig:Target} and \ref{Fig:Knowledge} present the results of this simulation study showing the performance metrics in terms of target collection (mission performance) and knowledge growth (collective intelligence), respectively.
%and BL2 has 50 M-robots. 


\section{Conclusion}
The results from the simulations indicate that the longer the agents retained the knowledge, the better the groups' performance was, with the assumption that no memory constraints were imposed on the agents. 
However, no significant difference in performance was observed when memory was constrained, except for when the memory was too low. This was due to the agent's inability to accommodate the new knowledge required to handle a new scenario. Further analyses are needed to gain an in-depth understanding of the impact of the ephemeral nature of knowledge in the performance of multiagent and swarm systems.

In the future, we extend our study on ephemeral learning to estimate the efficiency of query-answer mechanisms and compare it against regular and indirect learning modes.

%%%%%%%%%%%%%%%%% BIBLIOGRAPHY IN THE LaTeX file !!!!! %%%%%%%%%%%%%%%%%%%%%%
%\vspace{-5pt}
\bibliographystyle{plain}
\bibliography{references,newrefs}

\end{document}